%% file: main.tex
\documentclass[10pt, conference, anonymous]{IEEEtran}
\IEEEoverridecommandlockouts
\usepackage{cite}
\usepackage{amsmath,amssymb,amsfonts}

\usepackage{graphicx}
\usepackage{textcomp}
\usepackage{xcolor}
\usepackage{xspace}
\usepackage{bm}
\usepackage{booktabs}
\def\BibTeX{{\rm B\kern-.05em{\sc i\kern-.025em b}\kern-.08em
    T\kern-.1667em\lower.7ex\hbox{E}\kern-.125emX}}
\usepackage{newunicodechar}
\usepackage{pgfplots}
\pgfplotsset{compat=1.17}
\usepackage{multirow}
\usetikzlibrary{patterns}
\usepackage{subcaption} 
\usepackage{tabularx}
\usepackage{booktabs}
\usepackage{float}
\renewcommand*\thetable{\Roman{table}}
\renewcommand*{\tablename}{TABLE}
\usepackage{cleveref}
\usepackage{algorithm}
\usepackage{bm}
\newunicodechar{≤}{\ensuremath{\leq}}

    \makeatletter
\def\endthebibliography{%
  \def\@noitemerr{\@latex@warning{Empty `thebibliography' environment}}%
  \endlist
}
\makeatother
\usepackage{booktabs}
\usepackage{amsmath}
\usepackage{amsfonts}
\usepackage{mathtools}
\usepackage{amsthm}
\usepackage{bm}
\usepackage{subcaption}
\usepackage{xspace}
\usepackage{algorithm}
\usepackage{algorithmic}

\makeatletter
\def\BState{\State\hskip-\ALG@thistlm}
\makeatother

\makeatletter
\newcommand{\linebreakand}{%
  \end{@IEEEauthorhalign}
  \hfill\mbox{}\par
  \mbox{}\hfill\begin{@IEEEauthorhalign}
}
\makeatother

\newtheorem{lemma}{Lemma}

\newcommand{\name}{AccMER\xspace}
\newcommand{\bu}{\mathbf{u}} 
\newcommand{\bpi}{\bm{\pi}} 
\newcommand{\eqdef}{\stackrel{\mathsf{def}}{=}}

\newcommand{\btau}{\bm{\tau}} 

\makeatother

\author{
  \IEEEauthorblockN{Kailash Gogineni*}
  \IEEEauthorblockA{\textit{The George Washington University} \\
    Washington, DC, USA \\
    kailashg26@gwu.edu}
  \and
  \IEEEauthorblockN{Yongsheng Mei*}
  \IEEEauthorblockA{\textit{The George Washington University} \\
    Washington, DC, USA \\
    ysmei@gwu.edu}
  \and
  \IEEEauthorblockN{Tian Lan}
  \IEEEauthorblockA{\textit{The George Washington University} \\
    Washington, DC, USA \\
    tlan@gwu.edu}
  \linebreakand 
  \IEEEauthorblockN{Peng Wei}
  \IEEEauthorblockA{\textit{The George Washington University} \\
    Washington, DC, USA \\
    pwei@gwu.edu}
  \and
  \IEEEauthorblockN{Guru Venkataramani}
  \IEEEauthorblockA{\textit{The George Washington University} \\
    Washington, DC, USA \\
    guruv@gwu.edu}
}

\begin{document}

\title{AccMER: Accelerating Multi-Agent Experience Replay with Cache Locality-aware Prioritization}

\maketitle
\pagestyle{empty}  
\begingroup\renewcommand\thefootnote{}
\footnotetext{*These authors contributed equally to this work.}
\endgroup

\begin{abstract}
Multi-Agent Experience Replay (MER) is a key component of off-policy reinforcement learning~(RL) algorithms. By remembering and reusing experiences from the past, experience replay significantly improves the stability of RL algorithms and their learning efficiency. In many scenarios, multiple agents interact in a shared environment during online training under centralized training and decentralized execution~(CTDE) paradigm. Current multi-agent reinforcement learning~(MARL) algorithms consider experience replay with uniform sampling or based on priority weights to improve transition data sample efficiency in the sampling phase. However, moving transition data histories for each agent through the processor memory hierarchy is a performance limiter. Also, as the agents' transitions continuously renew every iteration, the finite cache capacity results in increased cache misses.

To this end, we propose \name, that repeatedly reuses the transitions~(experiences) for a window of $n$ steps in order to improve the cache locality and minimize the transition data movement, instead of sampling new transitions at each step. Specifically, our optimization uses priority weights to select the transitions so that only high-priority transitions will be reused frequently, thereby improving the cache performance. Our experimental results on the Predator-Prey environment demonstrate the effectiveness of reusing the essential transitions based on the priority weights, where we observe an end-to-end training time reduction of $25.4\%$~(for $32$ agents) compared to existing prioritized MER algorithms without notable degradation in the mean reward.

\end{abstract}


\input{Introduction}

\input{Motivation}
\input{Background}
\input{Methodology}

\input{Evaluation}
\input{Relatedwork}

\input{conclusion}

\section*{Acknowledgment}
This research is based on work supported by the National Science Foundation under grant CCF-2114415.

\bibliographystyle{IEEEtran}
\bibliography{references}





\end{document}

%% file: Introduction.tex
\section{Introduction}
\label{Introduction}
Reinforcement Learning~(RL) has been applied to solve many single-agent sequential decision-making problems~\cite{sutton2018reinforcement}. RL frameworks optimize the control of agent behavior and its interactions with the environment by taking actions based on current observation/state space, assessing the quality of state-action pairs using a reward function, and then transitioning to a new state~\cite{sutton2018reinforcement}. The function that determines the action is known as a policy. The agent strives to find the optimal policy to maximize the total cumulative~(discounted) reward. The function representing the reward estimates is known as the value function. 

Often times in practice, RL tasks involve multiple agents sharing the same environment, e.g., autonomous driving~\cite{cao2012overview, hu2019interaction}, robotics and planning~\cite{matignon2012coordinated, levine2016end}, and aviation systems~\cite{razzaghi2022survey}. Multi-agent reinforcement learning (MARL)~\cite{sutton2018reinforcement} 
helps to coordinate the decision-making among multiple agents and learn the desired joint behavior from collective experiences~(transition data) and achieve their goals. In particular, joint actions among these agents could affect the environment dynamically. The transitions observed in the environment are usually stored as experience tuples in a memory replay buffer and repeatedly used to improve the sample efficiency and policy training. This phase in the MARL training is called \textit{mini-batch sampling}. 
We note that the mini-batch sampling is compute-intensive in multi-agent systems, with each agent collecting a significant number of experience tuples of all other agents in every iteration to share information amongst the agents for collective decision making~\cite{gogineni2023scalability}. Consequently, the computational and memory bandwidth demands also increase exponentially with the number of agents, which limits the applicability of MARL in real-world decision-making situations~\cite {iqbal2019actor}.


Prior studies on experience replay buffers in RL have proposed various strategies to improve the transition data sampling efficiency. The simplest and most widely used experience replay method is uniform sampling, where the transition data stored in the replay buffer are sampled uniformly at random~\cite{lowe2017multi}. However, uniform sampling might often select unimportant transitions and slow down the learning efficiency. For this reason, prioritized experience replay~(PER)~\cite{schaul2015prioritized} and its variants were introduced~\cite{brittain2019prioritized,liu2021regret}.~However, most of these prior efforts focus on prioritization methods for the experience replay in single-agent settings, and they cannot be adapted readily to MARL scenarios. 

Recent work~\cite{mei2023mac} on collective priority optimization in Multi-Agent Experience Replay (MER) showed rigorous theoretical analysis in assigning optimal sampling weights to achieve higher mean rewards. 
MAC-PO prioritization technique achieves better convergence than the existing multi-agent learning algorithms. 
Our preliminary experiments (Section~\ref{motivation}) show that this implementation would still be computationally expensive as MAC-PO has to move the transition data histories and update its transitions in every iteration, resulting in cache and memory bandwidth bottlenecks. 
Thus, realizing efficient MARL algorithms with prioritization schemes from the systems perspective is still an open research problem.


In this paper, we propose \name, a cache-aware transition data reuse strategy to improve the MER efficiency for MARL algorithms. Specifically, we design transition data-reuse optimization that improves the cache locality by efficiently reusing higher-priority transitions during the MARL training phase. 
To the best of our knowledge, this is the {\it first} work to focus on improving the end-to-end training time of cooperative MARL algorithms. We validate the effectiveness of \name on the Predator-Prey environment~\cite{bohmer2020deep} through comparison with baseline MARL settings~(QMIX, WQMIX~\cite{rashid2020weighted}, and QPLEX), decomposed policy gradient method (i.e., VDAC~\cite{su2021value}). In our experiments, \name achieves an end-to-end training time reduction of $25.4\%$~(for $32$ agents) compared to MAC-PO without any significant degradation in the mean reward. 

The main contributions of our paper are the following:
\begin{itemize}
    \item We present \name, an experience data-reuse strategy that can be used in conjunction with MER to address MARL performance bottlenecks. In particular, we use experience prioritization to reuse high-priority transitions for future sampling.
    \item We adopt a hardware-software co-design approach where a cache-aware transition data-reuse strategy significantly reduces the last-level cache misses while ensuring the convergence levels to be on par with the best-performing state-of-the-art multi-agent algorithms.
    \item Our experimental results on the Nvidia Ampere systems demonstrate that: $1)$ \name reduces the end-to-end training time by about $25.4\%$~(for $32$ agents) on a cooperative multi-agent setting~(Predator-Prey task with no punishment); and $2)$ Interestingly, as the number of agents increases, our optimization efforts lead to a better convergence while reducing the training time on the same hardware compared to the current multi-agent prioritized experience replay~\cite{mei2023mac}.
\end{itemize}

%% file: Motivation.tex
\section{Motivation}
\label{motivation}
In this section we motivate the need to understand the performance bottlenecks in MARL algorithms, where we profile various state-of-the-art MARL frameworks implemented using actor-critic methods with usually very large state spaces. 

\begin{figure}[h]
  \includegraphics[width=0.48\textwidth]{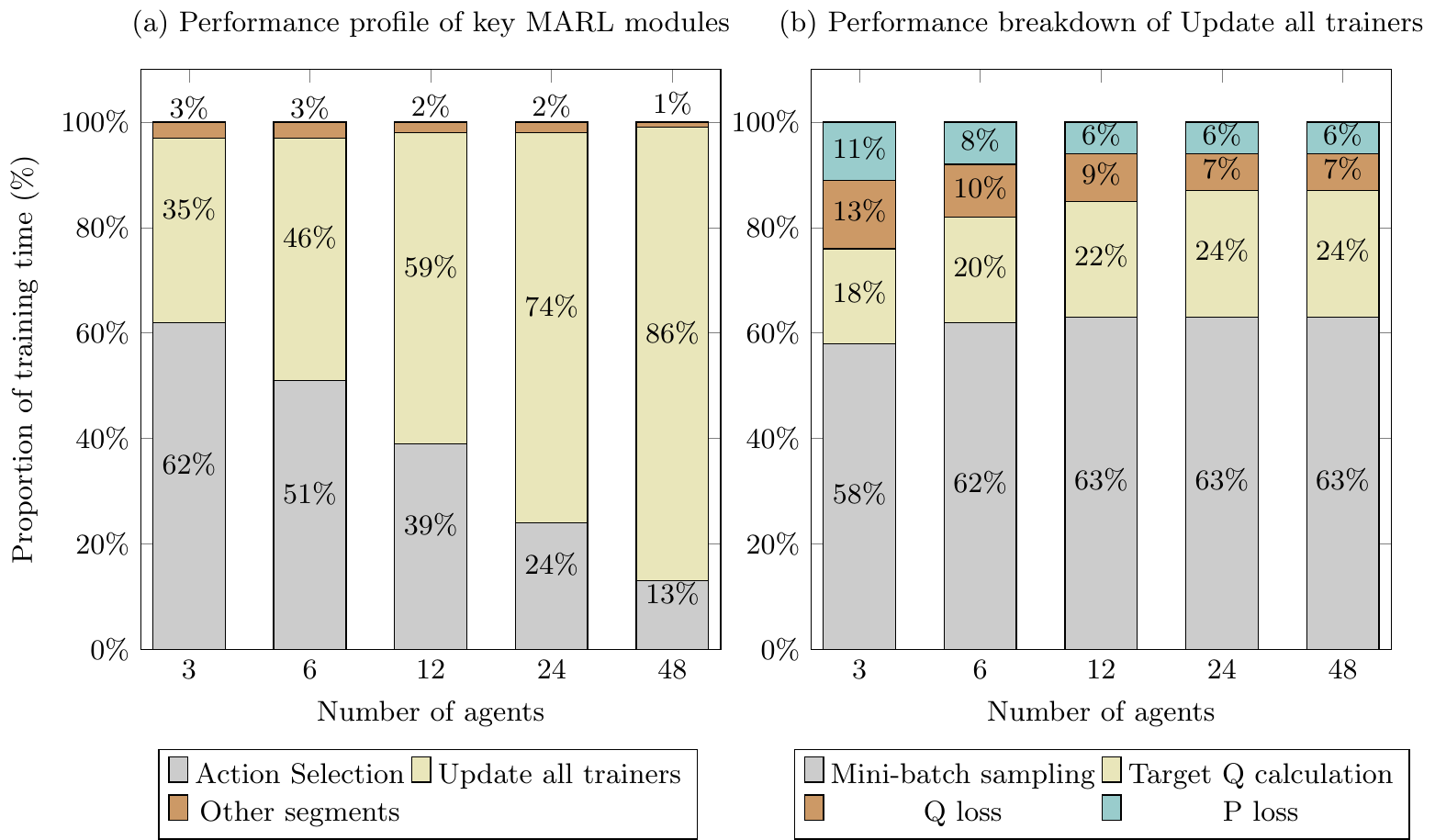}
    \caption{Training time breakdown of various functions averaged acorss several MARL workloads under multi-agent settings on Ampere Architecture~RTX 3090. The simulated multi-agent particle environment is Predator-Prey~\cite{lowe2017multi}.}
    \label{Figure1}
\end{figure}


We characterize the training phases of various representative MARL algorithms, including MADDPG~\cite{lowe2017multi}, MATD3~\cite{ackermann2019reducing}, and MASAC~\cite{ackermann2019reducing} using Predator-Prey environment~\cite{lowe2017multi}. 
These require multi-agent settings and learn the desired joint behavior from the collective experiences. 
These approaches employ memory replay buffers with a uniform sampling of transition history. 

Figure~\ref{Figure1}a illustrates how the proportion of training time changes when the number of agents increases linearly on a CPU-GPU platform. We note that the \textit{Update all trainers} phase contributes to $\approx$35\% to $\approx$86\% of the training time as the number of MARL agents grow from $3$ to $48$. The \textit{Action selection} phase involves individual agents' policy networks using local observations to interact with the environment and this phase scales linearly. During \textit{Update all trainers} phase, the actor network is updated using Q-values computed by the critic~\cite{lowe2017multi}. The target networks are created to achieve training stability. Note that the updating sequence of networks in the \textit{Update all trainers} phase begin with that of critics, followed by the actors, and then the target networks. \textit{Other segments} is a combination of reward collection, storing the present experiences and policy initialization and put together, they add a negligible performance overhead. 

For a deeper analysis, as shown in Figure~\ref{Figure1}b, we divide the \textit{Update all trainers} into multiple modules:~\textit{Mini-batch sampling, Target Q calculation}, and \textit{Q loss \& P loss} and present our results in the Predator-Prey environment. We observe that mini-batch sampling phase dominates the \textit{Update all trainers} as it occupies $60\%$ of training time in the update phase. This is because, each agent has to randomly sample a mini-batch of transition data~(history) of all other agents from the replay buffer to update both the critic and actor networks. In real-world systems, this task may be lead to huge compute and memory requirements as the number of agents increase. 

We also perform experiments to understand how the cache performance is affected as a consequence of batch size vs. buffer size trade-offs on the QMIX algorithm~\cite{rashid2018qmix} on account of the random memory access patterns. From Table~\ref{tradeoff}, when the batch size increases from $64$ to $256$ in the difficulty-enhanced predator-prey environment, the LLC load misses increase to $120\%$ in QMIX~(for $8$ agents). This is because of the uniform sampling, where the collected transitions will be continually renewed at every step, and finite cache capacity results in increased cache misses. As a result, conventional sampling is impractical and may lead to computing bottlenecks in real-world systems, especially when the number of agents scale under MARL. Furthermore, even after MER prioritization is enabled, the global cache misses grow by $2.5\times$ when the number of agents scale from $16$ to $32$ in MAC-PO~\cite{mei2023mac}. The dynamic memory requirements of observation and action spaces also grow quadratically due to each agent coordinating with other agents toward sharing their observations and actions~\cite{gogineni2023scalability, sheikh2020multi}.
\begin{table}[ht]
	\centering
    \caption{Cache miss profiles in the QMIX algorithm for different mini-batch and experience replay buffer sizes}
	\label{tradeoff}
	\begin{tabular}{lccc}
		\toprule
		Buffer size & Batch size & LLC load misses & global cache misses \\
		\midrule 
		100,000 & 256 & 10,228,039,764 & 30,727,770,917\\
		10,000 & 128 & 7,253,378,442 & 24,608,100,772\\
        1000 & 64 & 4,695,948,584 & 17,722,615,186\\
        100 & 16 & 2,247,388,137 & 9,428,462,486\\
		\bottomrule
	\end{tabular}
\end{table} 

%% file: Background.tex
\section{Background}
\label{background}
In this work, we consider a multi-agent sequential decision-making task as a decentralized partially observable Markov decision process (Dec-POMDP)~\cite{oliehoek2016concise} consisting of a tuple $ G=\langle S,U,P,R,Z,O,n,\gamma \rangle $, where $ s \in S $ describes the global state of the environment. At each time-step, each agent $ a \in A \equiv \{ 1,\dots,n \} $ selects an action $ u_a \in U $, and all selected actions combine and form a joint action $ \mathbf{u} \in \mathbf{U} \equiv U^n $. Such a process leads to a transition in the environment based on the state transition function $ P(s'|s,\mathbf{u}):S \times \mathbf{U} \times S \rightarrow [0,1] $. All agents share the same reward function $ r(s,\mathbf{u}):S \times \mathbf{U} \rightarrow \mathbb{R} $ with a discount factor $ \gamma \in [0,1) $.

In the partially observable environment, the agents' individual observations $ z \in Z $ are generated by the observation function $ O(s,u):S \times A \rightarrow Z $. Each agent has an action-observation history $ \tau_a \in T \equiv (Z \times U)^*$. Conditioning on the history, the policy becomes $ \pi^a(u_a|\tau_a):T \times U \rightarrow [0,1] $. The joint policy $ \bpi $ has a joint action-value function: $ Q^\pi(s_t, \mathbf{u}_t)=\mathbb{E}_{s_{t+1:\infty},\mathbf{u}_{t+1:\infty}}[R_t|s_t,\mathbf{u}_t] $, where $ t $ is the timestep and $R_t=\sum_{i=0}^{\infty} \gamma^i r_{t+i}$ is the discounted return. The learning algorithm has access to all local action-observation histories $\btau$ and global state $ s $ during training, yet every agent can only access its individual history in execution. The learning algorithm we use in this work is an actor-critic method, MAC-PO, a multi-agent prioritized experience replay variant of QMIX.

In QMIX~\cite{rashid2018qmix}, the learner is designed for multi-agent cooperative tasks with global reward. Specifically, QMIX solves credit assignment problems using additional information during training. The core idea of QMIX is value decomposition. A parameterized mixer function is proposed to combine the Q-functions of agents into a centralized Q-function that is trained on the global reward. Further, QMIX demonstrates that the mixer network provides monotonicity in its inputs, which makes the argmax of the agents' Q-functions consistent with the argmax of the centralized Q-function. This property is called Individual-Global-Max~(IGM), and it is important for factorizing the global Q-function to agents' Q-functions. 

%% file: Methodology.tex
\section{Methodology}
\label{methodology}
\name trains with the cache-aware transition data-reuse optimization on top of the MER prioritization scheme to improve the MARL performance. \name aims to reduce the number of last-level cache misses by reusing transition data, that ultimately improves the training time. The rest of this section delves deeper into the MER prioritization scheme and the transition data-reuse optimization of \name's design.

\subsection{Prioritization Optimization for Experience Replay}

Recent work~\cite{fujimoto2020equivalence} shows that the design of prioritized sampling methods influences the loss function. On the contrary, the expected gradient of a loss function with non-uniform sampling is equivalent to that of a weighted loss function with uniform sampling, which provides a recipe for transforming a regular loss function $ L_1 $ with a non-uniform sampling scheme into an equivalent weighted loss function $ L_2 $ with uniform sampling. Based on this equivalence, MAC-PO~\cite{mei2023mac} further explores the optimal weighting scheme for prioritized experience in MARL, given by a weighting factor $ w_k(s,\bu) $ when computing the loss function. In this paper, we adopt this MER prioritization scheme as the baseline while investigating the hardware-aware optimization that can improve learning efficiency regarding the overall training time and cache usage. Given the prioritization weight $w_k$, 
we use the following loss function during the learning, which is:
\begin{equation}
    L_{\rm \name} = \sum_{i=1}^{b}w_k(s,\bu)(Q_k-y_i)^2(s,\bu),
    \label{eq:macpo_loss}
\end{equation}
where $ b $ is the batch size. In the loss function~\eqref{eq:macpo_loss}, $ y_i=\mathcal{B}^*Q_{k-1} $ denotes a fixed target that can be obtained through a target network, where $\mathcal{B}^*$ is the Bellman operator satisfying $ \mathcal{B}^*Q(s,\bu) \eqdef r(s,\bu) + \gamma\arg\max_{\bu'}\mathbb{E}_{s'}Q(s',\bu') $. The following lemma for deciding optimal weights is proposed in~\cite{mei2023mac}.
\begin{lemma}[Optimal prioritization weight]
\label{lemma1}
The optimal weight in~\eqref{eq:macpo_loss} is proportional to:
\begin{equation}
    w_k(s,\bu)\propto|Q_k-\mathcal{B}^*Q_{k-1}|\exp(-|Q_k-Q^*|)f(\pi^a_k),
    \label{eq:w_k}
\end{equation}
where $ Q^* $ denotes the optimal action value function and the function $ f(\cdot) $ is defined as:
\begin{equation}
    f(\pi^a_k) \eqdef 1+\sum_{i=1}^{n}\prod_{\substack{j=1 \\ j \neq i}}^{n}\pi^j_k-n\prod_{i=1}^{n}\pi^i_k.
    \label{eq:f_pi}
\end{equation}
\end{lemma}
The optimal weight in~\eqref{eq:w_k} consists of three main terms, which are Bellman error term $ |Q_k-\mathcal{B}^*Q_{k-1}| $, value enhancement term $ \exp(-|Q_k-Q^*|) $, and joint action probability function $ f(\pi^a_k) $. The Bellman error term measures the distance between the estimation of the action value function and the Bellman target, in which the significant difference means higher hindsight Bellman error and will lead to higher sampling weight assignment. The value enhancement term indicates that any transitions with more accurate action values compared to the optimal value estimation after the Bellman update should be assigned with higher weights. Considering the relationship between agents' individual policies, the joint action probability function shows the counter-intuitive fact that the higher weights will be assigned to transitions with one's action differentiated from the others, as the maxima of function~\eqref{eq:f_pi} can be reached if and only if one agent's action probability is small in the transition while all other agents' action probabilities are large. We adopt this optimal weighting scheme in our work with the necessary normalization. To thoroughly exploit transitions with higher weights, we propose \name, that leverages data reuse strategy to efficiently remember and recapture the highly-weighed transitions in the replay buffer and utilize them for a specific number of steps for more efficient and performance-wise better learning.

\begin{algorithm}
	\caption{\name}
	\label{alg:AccMER}
	\begin{algorithmic}[1]
        \STATE Initialize step $t$, experience replay buffer $ \mathcal{D} $ with size $d$, mini-batch size $b$, reuse ratio $\alpha$, and weights
		\FOR{$ t =1:t_{max} $ }
        \STATE Initialize the state~(observation vetor)
		\WHILE{goal state is not reached}
        \STATE for each agent $a$, select action $u_a$ w.r.t. the policy $\pi^a$
    	\STATE Compute the reward and next state
    	\STATE Store the current trajectory (current state, next state, rewards, action) into replay buffer $ \mathcal{D} $
		\ENDWHILE
        \STATE for every $\lfloor d/b \rfloor$ steps, select the $\mathcal{S}^-$ = \{$\alpha \cdot b\}$ transitions from $ \mathcal{D}$ ranked based on the optimal weights 
        \STATE Sample $(1-\alpha) \cdot b$ transitions as $\mathcal{S}^+$ from the complement of $\mathcal{S}^-$ in $\mathcal{D}$ following the uniform distribution
        \STATE Update the mini-batch $\mathcal{S} = \mathcal{S}^- \cup \mathcal{S}^+$
		\FOR{each time-step $ k $ in $ \mathcal{S} $}
        \STATE Compute the optimal weights $w_k$ according to the prioritization scheme in Lemma~\ref{lemma1}
        \STATE Update the weights for transitions
		\ENDFOR
		\STATE Update the network parameters
		\ENDFOR
	\end{algorithmic}
\end{algorithm}

\subsection{Cache-aware transition data-reuse optimization}
\begin{figure}[ht]
  \includegraphics[width=0.5\textwidth]{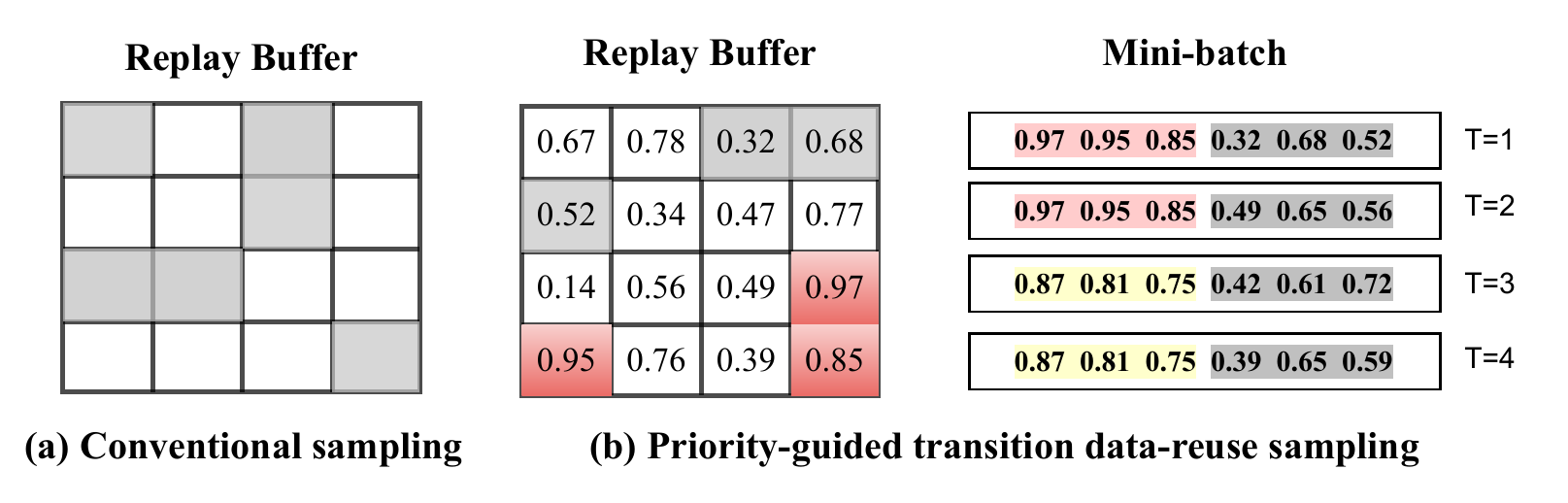}
    \caption{Illustration of (a) conventional sampling, where gray filled boxes denote the uniformly sampled transitions from the replay buffer, and (b)~data reuse sampling, we sample $50\%$ and reuse the rest of the transitions~(reuse ratio = $0.5$, batch size $b=6$, replay buffer $\mathcal{D}$ is a $4*4$ matrix) and the number of reuses~(time-steps) $n$ is computed by $\lfloor d/b \rfloor=2$. At step $T=1$, as the data reuse ratio is $0.5$, \name selects the first three transitions with the highest weights, and in the next phase, uniform sampling is performed on the remaining three transitions. At step $T=2$,~the same transition data will be reused, whereas a new set of transitions is sampled randomly. Since $n=2$, for $T=3$ and $T=4$, the reuse-based transition data updates every two steps.}
    \label{Figure2}
\end{figure}

From our analysis, we note that the sampling phase is one of the compute-intensive phases, as each agent has to sample all other agents' transition data sequentially. Figure~\ref{Figure2} shows an example layout of conventional sampling and the priority-guided transition data-reuse optimization.

Uniform sampling suffers from random memory access patterns, often leading to low cache line utilization, meaning the arrays' transitions are indexed randomly. Because of this, the number of cache misses and memory bandwidth demands of the program increase with the number of agents.
 
To perform the cache-aware transition data-reuse optimization, \name first partitions two different micro-batches according to the mini-batch size~$b$ as shown in Algorithm~\ref{alg:AccMER}. We initialize a weight lookup table $\mathcal{W}$ mapping to transitions data addresses in the replay buffer $ \mathcal{D}$. Both lookup table $\mathcal{W}$ and replay buffer $\mathcal{D}$ have the size of $d$. The initial weights for the weight lookup table will be the same for all transitions. Depending on the reuse ratio $\alpha \in [0,1]$~(if $\alpha$ is 0, all the transitions are sampled uniformly, where as if $\alpha$ is 1, all the transitions will be reused), the micro-batch $\mathcal{S}^-$ ranks and selects the transitions data according to the weight lookup table $\mathcal{W}$ from the replay buffer $ \mathcal{D}$~(line 9). 
For every $\lfloor d/b \rfloor$ steps, we reuse the same transitions according to the priority weights. In this way, we map the addresses of the weight lookup table $\mathcal{W}$ and transitions in the replay buffer $\mathcal{D}$ to choose high-priority transitions. Unlike conventional sampling, which might select random transitions and often select unimportant transitions, our optimization leverages priority-guided transition data-reuse to improve the data availability in upper-level caches for better memory locality.

Note that reuse ratio, $\alpha$ for the transition data partition is input by the user. We sample the remaining transitions from the complement of $\mathcal{S}^-$ in replay buffer~$\mathcal{D}$ following the uniform distribution and store it in $\mathcal{S}^+$~(line 10). In the next step, we concatenate the $\mathcal{S}^-$ and $\mathcal{S}^+$ and update the batch $\mathcal{S}$~(line 11).~By Lemma~\ref{lemma1},~we calculate the optimal prioritization weights for the transitions, and then we update the weight lookup table $\mathcal{W}$ with the optimal weight $w_k$. In order to avoid the bias towards reusing the same transitions, we apply a discount factor~$\gamma$ to all the weights in the weight lookup table $\mathcal{W}$. To maximize the reuse of transitions with high priority, we sort the weight lookup table $\mathcal{W}$ so that high-priority transitions will be reused for multiple steps.

%% file: Evaluation.tex
\section{Experimental Evaluation}
\label{experimental-evaluation}
In this section, we evaluate the performance of \name using difficulty-enhanced Predator-Prey task~\cite{mei2023mac} at various punishment levels. Section~\ref{evaluation-setup} introduces the implementation details and hyper-parameters that we used in the experimental evaluation. In Section~\ref{subsection2}, we demonstrate the effectiveness of \name by comparing the mean reward with several state-of-the-art MARL baselines. In Section~\ref{subsection2}, we show that \name is able to improve the end-to-end training time and reduce a significant number of last-level cache misses with no significant loss in the mean episode reward. Finally, in Section~\ref{subsection4}, we conduct the additional scalability tests to demonstrate the benefits of \name\footnote{https://github.com/kailashg26/AccMER}.

\subsection{Evaluation Setup}
\label{evaluation-setup}
For evaluation, we implemented the key components of \name: a) Optimal prioritization~\cite{mei2023mac}, and b) Cache-aware transition data-reuse optimization on the baseline QMIX code base. We use epsilon greedy for action selection with annealing from $ \epsilon $ = 0.995 decreasing to $ \epsilon $ = 0.05 in 100K training steps in a linear way~\cite{sutton2018reinforcement, mei2023mac}. The performance for each algorithm is evaluated for 32 episodes every 1000 training steps. 
\begin{table}[h]
	\centering
	\caption{Hyper-parameters for the Predator-Prey task with no punishment.}
	\label{tab:hyper}
	\begin{tabular}{lc}
		\toprule
		Hyper-parameter & Value \\
		\midrule 
		Batch size & 256 \\
		Replay buffer size & 100000 \\
		Target network update interval & Every 200 episodes \\
		Learning rate & 0.001 \\
		TD-lambda & 0.6 \\
		\bottomrule
	\end{tabular}
\end{table}

\textit{Target Platform}:
\textcolor{black}{We train and profile \name on the Nvidia GeForce~RTX~2080TI architecture connected with  Intel(R) Core(TM) i9-7920X CPU, which has 12 cores with 16 MiB of Last-Level Cache, 128 GiB Gigabytes of main memory and the CPU's clock speed of 2.90GHz. The server runs on Ubuntu Linux~20.04.5~LTS operating system with CUDA~11.3, cuDNN~8.2, PCIe Express®~v3.0 with NCCL~v2.5.7 communication library. The machine supports python 3.8.16,~PyTorch~(v1.8.2), pyyaml~(v5.3.1) and OpenAI GYM~(v0.11). We use Perf~\cite{de2010new} tool to evaluate the hardware efficiency.}

\begin{table}[h]
	\centering
	\caption{Hyper-parameters for the Predator-Prey task with punishment$=-1.5$.}
	\label{tab:hyper1}
	\begin{tabular}{lc}
		\toprule
		Hyper-parameter & Value \\
		\midrule 
		Batch size & 128 \\
		Replay buffer size & 10000 \\
		Target network update interval & Every 200 episodes \\
		Learning rate & 0.001 \\
		TD-lambda & 0.6 \\
		\bottomrule
	\end{tabular}
\end{table}
\begin{figure*}[ht!]
	\centering
	\begin{subfigure}[H]{0.48\textwidth}
		\centering
		\includegraphics[width=\textwidth]{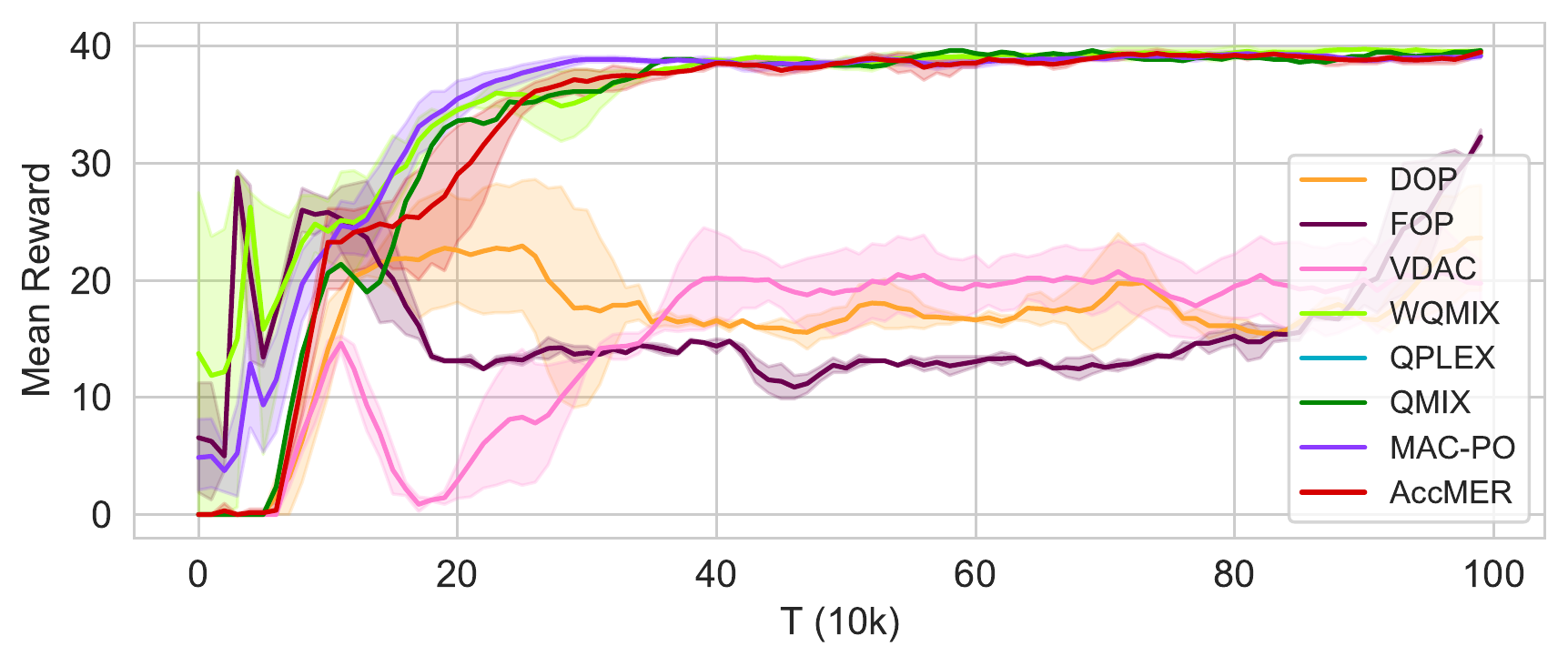}
		\caption{\small No punishment}
		\label{fig:pp_0}
	\end{subfigure}
	\begin{subfigure}[H]{0.50\textwidth}
		\centering
		\includegraphics[width=\textwidth]{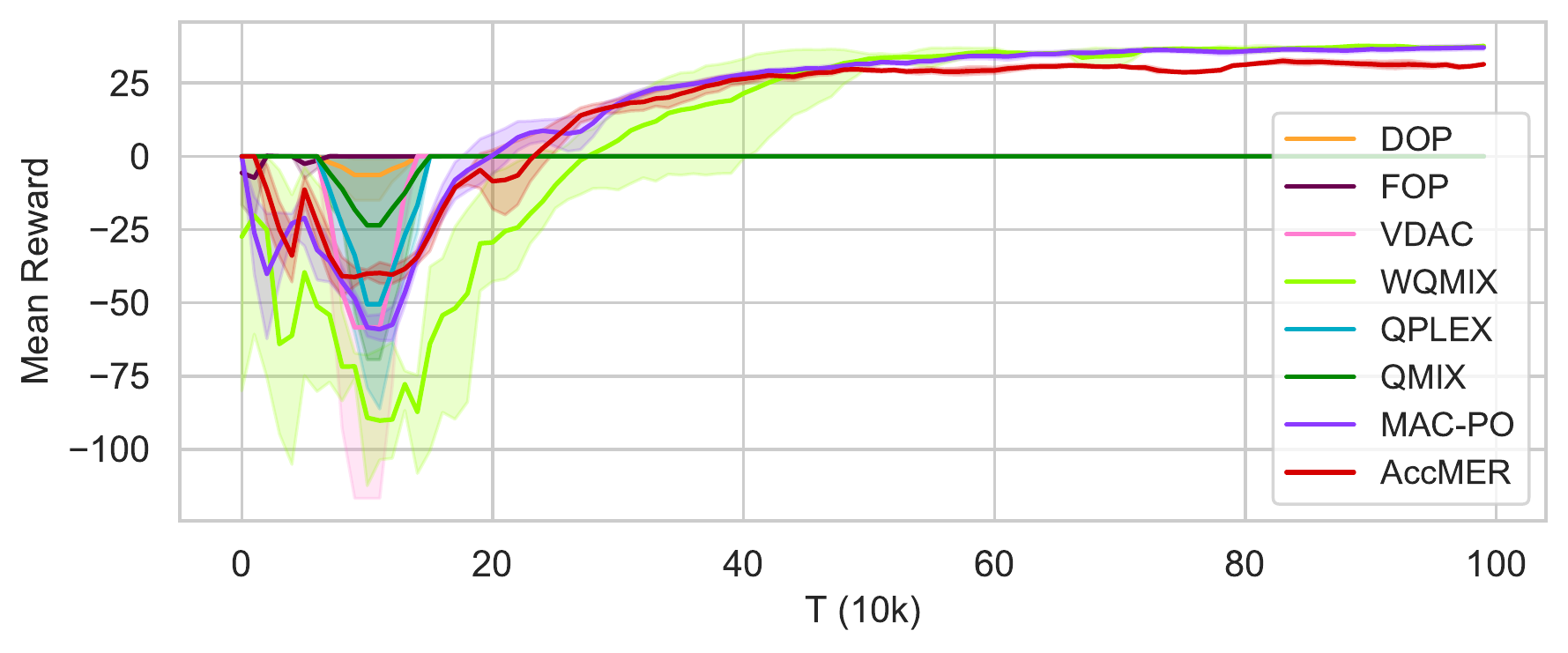}
		\caption{\small Punishment $ = - $1.5}
		\label{fig:pp_1.5}
	\end{subfigure}
	\caption{Average reward per episode on the Predator-Prey tasks for \name and other MARL algorithms under different punishment levels. \name shows almost the same convergence speed as MAC-PO while reducing the total training time.}
	\label{fig:pp}
\end{figure*}
\textit{Multi-agent environment}:
A partially observable environment on a  grid-world Predator-Prey task~\cite{mei2023mac} where 8 agents have to catch 8 prey in a 10 × 10 grid. Each agent can either move in one of the 4 compass directions, remain still, or try to
catch any adjacent prey. In this task, a successful capture with the positive reward of 1 must include two or more predator agents surrounding and catching the same prey simultaneously, requiring a high level of cooperation. A failed coordination between agents to capture the prey, which happens when only one predator catches the prey, will receive a negative punishment reward. We select the punishments of 0 and $-$1.5 in the experiments, with more punishment representing higher difficulty.
\begin{figure*}[ht!]
	\centering
	\begin{subfigure}[H]{0.48\textwidth}
		\centering
		\includegraphics[width=\textwidth]{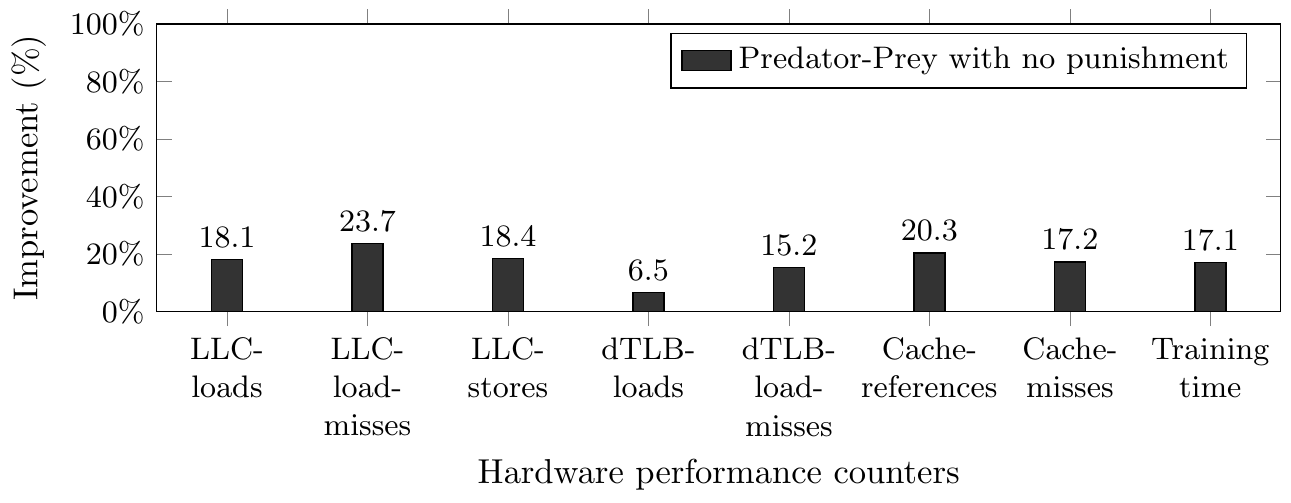}
		\caption{\small No punishment}
		\label{perfpp_0}
	\end{subfigure}
	\begin{subfigure}[H]{0.50\textwidth}
		\centering
		\includegraphics[width=\textwidth]{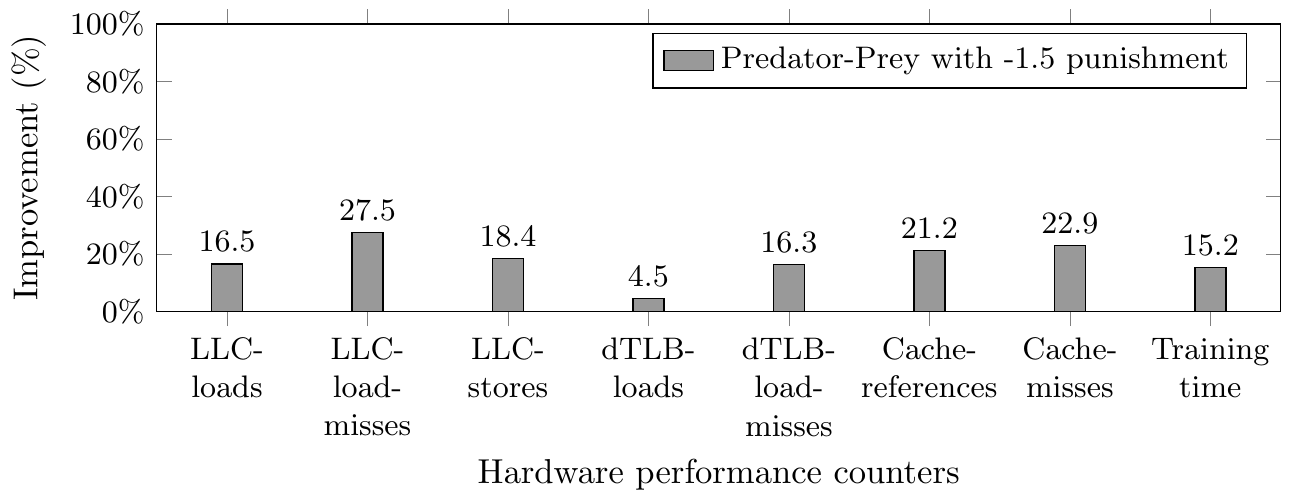}
		\caption{\small Punishment $ = - $1.5}
		\label{perfpp_1.5}
	\end{subfigure}
	\caption{Hardware performance analysis of \name for Predator-Prey environment with punishment=~$0$ and punishment=~$-1.5$.}
	\label{figperf}
\end{figure*}
\subsection{Comparison with MARL baselines}
\label{subsection2}
We select multiple state-of-the-art MARL algorithms for comparison, which include value-based factorization MARL algorithm (i.e., QMIX~\cite{rashid2018qmix}, WQMIX~\cite{rashid2020weighted}, and QPLEX), decomposed policy gradient method (i.e., VDAC~\cite{su2021value}), and decomposed actor-critic approaches (i.e., FOP~\cite{zhang2021fop} and DOP~\cite{wang2020dop}). All of the baselines have demonstrated their convergence properties in handling various multi-agent tasks.

We validate the efficacy of \name on Predator-Prey environment for two punishment levels:~$0$ and $-1.5$, and the hyper-parameters are shown in Table~\ref{tab:hyper} and Table~\ref{tab:hyper1}. The transition data-reuse ratio $\alpha$ for all the experiments is $0.5$, and the batch size $b$ for Predator-Prey:~$0$ and $-1.5$ is $256$ and $128$, respectively. Additionally, we set the discount factor $\gamma$ as $0.8$ for the Predator-Prey:~$-1.5$ hard settings and $\gamma$ as $1.0$ for Predator-Prey~(no punishment).

Figure~\ref{fig:pp} shows the performance of eight algorithms with different punishments, where all results show the effectiveness of \name. We note that \name's convergence levels are on par with best-performing state-of-the-art MARL algorithms in finding the optimal policy. In Figure~\ref{fig:pp_1.5}, \name significantly outperforms other state-of-the-art algorithms like QMIX and WQMIX and performs on par with MAC-PO in a hard setting that requires a higher level of coordination among agents in order to learn optimal policy. Most of the MARL algorithms learn a sub-optimal policy where agents learn to work together with limited coordination. Although the algorithmic performance~(reward) of \name and MAC-PO are almost similar, compared to the latter, Figure~\ref{figperf} shows that \name achieves a performance speedup by reducing the training time by about 17\% for Predator-Prey~(no punishment) task compared to MAC-PO. This demonstrates that efficiently recapturing the prioritized transitions with higher weights and smart cache data-reuse strategies, \name can learn the optimal policy and improve the training efficiency.

\subsection{Impact of \name} 
\label{subsection3}
We perform experiments with a transition data-reuse ratio, $\alpha=0.5$, meaning 50\% of the prioritized transitions are being reused between the iterations. 

Figure~\ref{figperf} shows the training time savings with respect to the wall-clock training time and the profiles of certain key hardware performance counters. Specifically, in a Predator-Prey environment with no punishment, compared to MAC-PO, \name reduces the end-to-end training time by about $17\%$, which is almost $1.2\times$ faster than MAC-PO~(Figure~\ref{perfpp_0}) and there is no noticeable degradation in the mean episode reward. In fact, by reusing prioritized transition data, we reduced the LLC-load misses by about $23.7\%$ and the global cache misses by about $17.2\%$. As the batch size is $128$ for the hard setting~(Predator-Prey with punishment=$-1.5$), \name shows $15.2\%$ improvement in training time, compared to the $17.1\%$  when the batch size is $256$. This indicates that larger batch sizes and replay buffers can boost further performance gains~(Figure~\ref{perfpp_1.5}). Interestingly, when a high level of coordination is required among the agents, data-reuse optimization demonstrates {\it higher} effectiveness by reducing $27.5\%$ LLC load misses, and $16.3\%$ dTLB load misses compared to MAC-PO. These experimental results confirm that priority-guided transition data-reuse is highly effective in multi-agent scenarios. We further note that excluding the environment interactions phase from the MARL training time will give further speedups, as the environment interactions grow dramatically when more agents are involved in the MARL training phase.

\subsection{Scalability Tests} 
\label{subsection4}
We conduct the scalability tests and profile \name on the Nvidia GeForce~RTX~3090 Ampere Architecture connected with AMD Ryzen Threadripper PRO 3975WX CPU, which has 32 cores with 128 MiB of Last-Level Cache, 512 Gigabytes of main memory and the CPU's clock speed of 3.5GHz. The server runs on Ubuntu Linux~20.04.5~LTS OS with PCIe Express®~v4.0 and NCCL~v2.8.4 communication library. 

We use the following hyper-parameters for the scalability tests: The transition data-reuse ratio $\alpha$ is $0.5$, and the batch size $b$ for Predator-Prey:~$0$ is set to $128$, with $10,000$ as the buffer size. Additionally, we set the discount factor $\gamma$ as $0.8$ for the Predator-Prey:~$0$ task.
\begin{figure}[ht]
  \includegraphics[width=0.5\textwidth]{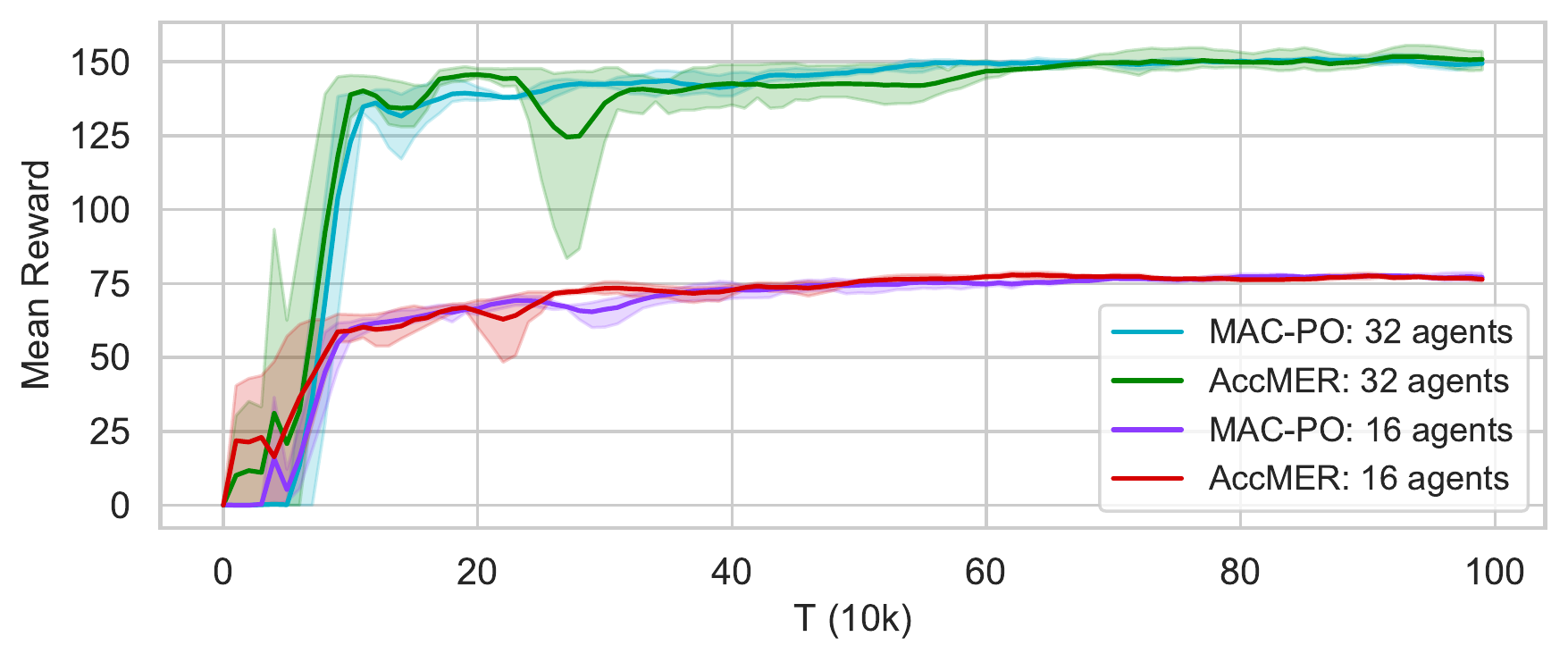}
    \caption{Average reward per episode on the Predator-Prey task for \name and MAC-PO algorithms with no punishment.} 
    \label{PP:16_32}
\end{figure}

Figure~\ref{PP:16_32} shows the mean episode reward curves of \name and MAC-PO~\cite{mei2023mac}. While we can see that when the number of agents increases linearly~(from $16$ to $32$), \name shows a slightly faster convergence than MAC-PO. Moreover, since \name reuses the prioritized transition data for multiple future steps, it reduces the end-to-end training time by about $19.8\%$ and $25.4\%$ for $16$ and $32$ agents, respectively~(Figure~\ref{perf_16_32}). By scaling the number of agents, we observe that the global cache misses gradually improve~($18.4\%$ improvement in cache misses for $32$ agents over the baseline), which indicates that \name can achieve higher speedups and improve the hardware efficiency for large-scale cooperative MARL settings. 



%% file: Relatedwork.tex
\section{Related Work}\label{relatedwork}
Prior works have demonstrated how experience replay and its variants can achieve better convergence for RL workloads. However, to our knowledge, no prior research presents insights into end-to-end performance improvement that involves multiple agents from the systems perspective. We provide an overview of related efforts in hardware-software acceleration and experience replay in RL. 
\begin{figure}[ht]
  \includegraphics[width=0.48\textwidth]{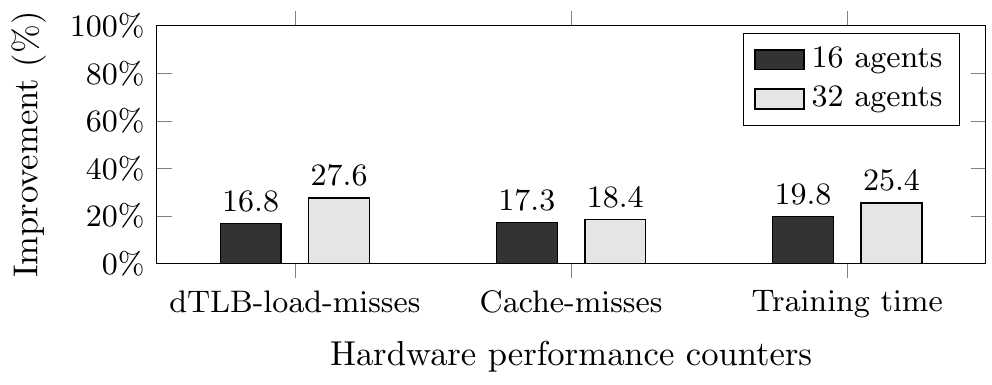}
    \caption{Performance analysis of \name for Predator-Prey task with no punishment in multi-agent settings.}
    \label{perf_16_32}
\end{figure}
\subsection{Quantization}
Low-precision training for neural networks reduces the neural network weights, enables faster compute operations, and minimizes the memory transfer computation time. Quantization aware training~\cite{hubara2017quantized}, post-quantization training~\cite{tambe2020algorithm}, and mixed precision~\cite{micikevicius2017mixed} demonstrated that neural networks may be quantized to a lower precision without significant degradation in rewards. Furthermore, to speedup the training, prior works have proposed algorithmic modifications~(e.g., compound loss scaling, storing the hypotenuse in Adam, etc.) that leaves the underlying agent and its hyper-parameters unchanged but improves the numerical stability and reduces the memory and compute requirements~\cite{bjorck2021low}. QuaRL~\cite{krishnan2022quarl} demonstrates how to accelerate single-agent RL, where quantization is applied to speedup the RL training and inference. 
All the prior works differ from our work as they apply quantization to single-agent RL algorithms or neural networks. In contrast, we seek to improve the performance of MARL by reusing the prioritized transitions for a certain number of time-steps.

\subsection{Software-Hardware Acceleration for RL} 
Distributed training has been widely adopted to reduce the training time of single-agent RL algorithms~\cite{babaeizadeh2016ga3c, cho2019fa3c, li2019accelerating, hoffman2020acme, stooke2018accelerated, clemente2017efficient}. Another strategy for MARL acceleration is to show the training efficiency via theoretical analysis by restricting the agent interactions to one-hop neighborhoods and adopting a distributed training strategy to simulate the state transitions of only a small subset of agents on each compute node~\cite{wang2022darl1n}. However, training on VM-based approaches still requires extensive management of the cluster and deploying the training jobs. 
FA3C~\cite{cho2019fa3c} studies how to accelerate multiple parallel worker scenarios, where each agent is controlled independently within their own environments using single-agent RL algorithms. 
iSwitch~\cite{li2019accelerating} reduces the end-to-end network latency for synchronous training, but also improves the convergence with faster weight updates for asynchronous training. 
In contrast, our work focuses on multi-agent learning frameworks, where the agents operate in a single shared environment. 

\subsection{Experience Replay Buffer}
Many RL algorithms adopt prioritization to increase the learning efficiency, initially originating from prioritized sweeping for value iteration~\cite{moore1993prioritized, van2013planning}. Prioritized experience replay~(PER)~\cite{schaul2016prioritized} is one of the key advancements in the DQN algorithm~\cite{van2016deep, wang2016dueling} and has been included in many RL algorithms combining multiple improvements~\cite{horgan2018distributed, barth2018distributed}. Variants of PER have been proposed for considering sequences of transitions~\cite{daley2019reconciling, brittain2019prioritized} or optimizing the prioritization function~\cite{zha2019experience}. 
Discor re-weights updates to reduce variance\cite{kumar2020discor}. ReMERN uses the regret minimization method to design the prioritized experience replay scheme in the single-agent environment\cite{liu2021regret}. So far, most prior works about experience replay are designed for single-agent RL algorithms. 

A recently proposed multi-agent experience replay framework, MAC-PO~\cite{mei2023mac}, can find the optimal prioritized sampling scheme by computing the optimal sampling weights for experience replay when the environment involves multiple agents. We adopt MAC-PO as one of the baselines in our studies. However, different from MAC-PO, this paper seeks to improve the actual run-time (system) performance of the multi-agent experience replay prioritization, such as training efficiency and cache utilization. Specifically, \name adopts the MER prioritization scheme in MAC-PO and repeatedly reuses the selected transitions with high prioritization weights, achieving performance speedup while retaining the MAC-PO's learning performance advantages. Apart from methods that focus on improving the training time of cooperative problems, other mechanisms use a \textit{neighbor sampling strategy} to improve the locality and training efficiency of competitive tasks~\cite{gogineni2023towards}.

%% file: conclusion.tex
\section{Conclusion}
\label{conclusion}
We presented \name, a cache-aware transition data reuse strategy for multi-agent experience replay. Our experimental results demonstrate that \name reduces the overall training time by about $25.4\%$~(for $32$ agents) over prior multi-agent prioritized replay schemes. Additionally, we show that the proposed data reuse optimization alleviates the performance issues posed by random memory access patterns by optimizing the transition data sampling phase for better hardware cache locality, thereby improving the overall MARL performance.